\documentclass[prd,amsmath,amssymb,floatfix,superscriptaddress,notitlepage,nofootinbib,preprintnumbers,twocolumn]{revtex4-1}
\usepackage{amsfonts,amssymb,amsmath,graphicx,color,bm,enumitem}
\usepackage[utf8]{inputenc}
\usepackage{amsthm}
\usepackage[english]{babel}
\definecolor{ultramarine}{rgb}{0.07, 0.04, 0.56}
\definecolor{cadmiumgreen}{rgb}{0.0, 0.42, 0.24}
\definecolor{indigo(dye)}{rgb}{0.0, 0.25, 0.42}
\usepackage[linktocpage=true]{hyperref}
\hypersetup{
colorlinks=true,
citecolor=blue,
linkcolor=red,
urlcolor=indigo(dye),
}

\newcommand{\be}{\begin{eqnarray}}  
\newcommand{\ee}{\end{eqnarray}}
\newcommand{\bem}{\begin{pmatrix}}
\newcommand{\eem}{\end{pmatrix}}

\newcommand{\tit}{\tilde{t}}
\newcommand{\tia}{\tilde{a}}
\newcommand{\F}{\mathcal{F}}
\newcommand{\G}{\mathcal{G}}

\newcommand{\Q}{\mathcal{Q}}
\newcommand{\R}{\mathcal{R}}

\newcommand{\W}{\mathcal{W}}
\newcommand{\D}{\mathcal{D}}
\newcommand{\A}{\mathcal{A}}
\newcommand{\B}{\mathcal{B}}
\newcommand{\C}{\mathcal{C}}
\newcommand{\X}{\mathcal{X}}
\newcommand{\Y}{\mathcal{Y}}
\newcommand{\Z}{\mathcal{Z}}

\begin{document}

\title{
Generalized disformal invariance of cosmological perturbations with second-order field derivatives
}

\author{Masato Minamitsuji}
\affiliation{Centro de Astrof\'{\i}sica e Gravita\c c\~ao  - CENTRA, Departamento de F\'{\i}sica, Instituto Superior T\'ecnico - IST, Universidade de Lisboa - UL, Av. Rovisco Pais 1, 1049-001 Lisboa, Portugal.}

\begin{abstract}
We investigate how the comoving curvature and tensor perturbations are transformed under the generalized disformal transformation with the second-order covariant derivatives of the scalar field, where the free functions depend on the fundamental elements constructed with the covariant derivatives of the scalar field with at most the quadratic order of the second-order covariant derivatives. Our analysis reveals that on the superhorizon scales the difference between the comoving curvature perturbations in the original and new frames is given by the combination of the time derivative of the comoving curvature perturbation, the intrinsic entropy perturbation of the scalar field, and its time derivative in the original frame. Thus,  in the case that on the superhorizon scales (1) the intrinsic entropy perturbation and its time derivative vanish and  (2) the comoving curvature perturbation in the original frame, $\R_c$, is conserved, the comoving curvature perturbation becomes invariant under the disformal transformation on the superhorizon scales. We also show that the tensor perturbations are also disformally invariant, in the case that the tensor perturbations in the original frame are conserved with time.
\end{abstract}
\date{\today}
\maketitle

\section{Introduction}
\label{sec1}

Cosmological inflation is recognized 
as the most promising scenario about the history in the early Universe
\cite{Linde:2005ht,Liddle:2000cg,Mukhanov:2005sc,Lyth:1998xn,Bassett:2005xm,Langlois:2010xc}.
Although the latest observational data of the large-scale anisotropies
of the Cosmic Microwave Background (CMB)
\cite{Planck:2013jfk,Ade:2015lrj,Aghanim:2018eyx,Akrami:2018odb}
strongly favour the simplest single-field and slow-roll inflation models,
models of inflation
driven
by various mechanisms,
e.g., 
multiple scalar fields,
other field species such as vector and spinor fields, 
modified kinetic terms,
nonminimal (derivative) couplings to the spacetime curvature,
and 
self-derivative interactions
have also been explored
\cite{kinf,dbiinf,Kobayashi:2010cm,Burrage:2010cu,ginf,Motohashi:2020wxj,Golovnev:2008cf,Kanno:2008gn,Bezrukov:2007ep,Germani:2010gm}.

For a long time,
the scalar-tensor theories with the higher-derivative interactions
have been thought to be problematic,
since the equations of motion for the metric and the scalar field
would generically contain derivatives higher than second-order,
indicating the appearance of the Ostrogradsky ghosts
\cite{Woodard:2015zca}.
Ref.~\cite{Zumalacarregui:2013pma}
argued that 
an invertible frame transformation
with the derivatives of the scalar field
maps a class of the conventional scalar-tensor theories 
to a new class without the Ostrogradsky ghosts,
despite the apparent higher-derivative features of the theory.
More explicit studies 
have revealed
that
the appearance of the Ostrogradsky ghosts
can be avoided
by imposing the certain degeneracy conditions amongst 
the equations of motion with the highest order time derivatives.
The scalar-tensor theories
under the imposition of the degeneracy conditions 
have been developed 
and 
are currently recognized
as the degenerate higher-order scalar-tensor (DHOST) theories 
\cite{Langlois:2015cwa,Achour:2016rkg,BenAchour:2016fzp,Langlois:2018dxi}.
The DHOST theories 
correspond to the most general scalar-tensor theories
with the single scalar field without the Ostrogradsky ghosts
and 
include all the previously known classes of the scalar-tensor theories,
especially, 
the Horndeski theories \cite{Horndeski:1974wa,Deffayet:2011gz,ginf}
and 
the beyond-Horndeski theories \cite{Gleyzes:2013ooa,Gleyzes:2014dya,Gleyzes:2014qga}.

In order to compare the inflationary models with the observational data,
we consider the linear perturbations about 
the Friedmann-Lem\^aitre-Robertson-Walker (FLRW) spacetime
\begin{eqnarray}
\label{scalar_flrw}
ds^2
&=&
g_{\mu\nu}
dx^\mu dx^\nu
\nonumber
\\
&=&
-\left(1+2A(t,x^i)\right)dt^2
+2a(t) \partial_i B(t,x^i) dt dx^i
\nonumber\\
&+&
a(t)^2
\left[
\left(1-2\psi (t,x^i)\right)\delta_{ij}
+2\partial_i \partial_j E (t,x^i)
\right.
\nonumber\\
&&
\left.
+
h_{ij} (t,x^i)
\right]
dx^i
dx^j,
\end{eqnarray}
where 
$t$ and $x^i (i=1,2,3)$ are the physical time and comoving spatial coordinates,
$a(t)$ is the cosmic scale factor, 
$A$, $B$, $\psi$, and $E$ are the scalar metric perturbation variables,
and $h_{ij}$ denotes the tensor metric perturbations 
obeying the transverse-traceless conditions $\delta^{ij}h_{ij}=0$ and $\partial^i h_{ij}=0$,
respectively.
We also consider the perturbation of the scalar field
\begin{eqnarray}
\label{decomp1}
\phi=\phi_0(t)+\phi_1(t,x^i),
\end{eqnarray}
ad  neglect the vector metric perturbations.
In order to see the dependence on the scales,
from now on, 
we decompose any perturbation variable $Q$ 
into the comoving Fourier modes 
\begin{eqnarray}
Q=
\int d^3k Q_k e^{ik_i x^i},
\end{eqnarray}
where $k_i$ is the comoving momentum vector and $k^2:=\delta^{ij} k_i k_j$,
although we will not show the subscript ``$k$'' explicitly.
After the Fourier transformation,
the spatial derivative $\partial_i$ and 
the Laplacian term $\Delta:=\delta^{ij}\partial_i\partial_j$ are replaced by 
$(-i k_i)$ and $(-k^2)$ in the perturbation equations,
respectively. 
Since in this paper we focus on the regime of the linearized perturbations,
there will be no coupling of the different $k$ modes.

While the tensor perturbations $h_{ij}$ are gauge-invariant,
the scalar perturbations $A$, $B$, $\psi$, $E$, and $\phi_1$
are not.
Thus, 
in order to compare with the observational data, 
we have to construct the
the gauge-invariant combinations of them
\cite{Bardeen:1980kt,Kodama:1985bj,Mukhanov:1990me}.
The gauge-invariant perturbations
relevant for the inflationary models
in the scalar-tensor theories  
are given 
by the combinations of the metric and scalar field perturbations.
The particularly important gauge-invariant quantity
is the comoving curvature perturbation
\cite{Bardeen:1983qw,Malik:2000ax,Lyth:2004gb,Gordon:2000hv}
\begin{eqnarray}
\label{com1}
\R_c
:=\psi 
+\frac{1}{\dot{\phi}_0}\frac{\dot{a}}{a}
\phi_1,
\end{eqnarray}
where 'dot' denotes the derivative with respect to the time $t$.
The spectral features of the comoving curvature perturbation 
are directly related to the data
of the large-scale CMB anisotropies.

In developing the new inflationary models 
in the more general scalar-tensor theories,
{\it frame invariance} of the cosmological obvservables 
should be very useful,
since it allows us to evaluate the observables 
in the frame which is technically the most convenient.
The conformal transformation
${\tilde g}_{\mu\nu}= \C(\phi)g_{\mu\nu}$,
where $\phi$ denotes the scalar field,
maps a class of the {\it conventional} scalar-tensor theories,
\be
\label{cano}
{\cal L}
=\xi (\phi)R -\omega(\phi)  \X -V(\phi),
\ee
with
the potential $V(\phi)$, 
the nonminimal coupling to the Ricci curvature $\xi(\phi)R$,
the kinetic function $\omega(\phi)$,
and the kinetic term of the scalar field
\begin{eqnarray}
\label{defx}
\X
&:=&
\phi^\nu \phi_\nu,
\end{eqnarray}
where we have defined the shorthand notation
for the covariant derivatives of the scalar field
by $\phi_{\mu\nu\cdots\alpha}:= \nabla_\alpha\cdots \nabla_{\nu}\nabla_{\mu}\phi$
and 
$\phi^{\mu\nu\cdots\alpha}:= \nabla^\alpha\cdots \nabla^{\nu}\nabla^{\mu}\phi
=g^{\rho\mu}g^{\sigma\nu} \cdots  g^{\alpha\beta}\nabla_\beta\cdots \nabla_{\sigma}\nabla_{\rho}\phi$
represent the covariant derivatives of the scalar field
associated with the metric $g_{\mu\nu}$, 
to another class
in which the structure of the Lagrangian density \eqref{cano}
is preserved
with the redefined functions of 
${\tilde \xi}(\phi)$, ${\tilde \omega}(\phi)$, and ${\tilde V} (\phi)$. 
Interestingly, 
it has been shown that
the gauge-invariant comoving curvature perturbation \eqref{com1}
is invariant
under the conformal transformation \cite{Makino:1991sg,Gong:2011qe,Chiba:2008ia},
which allows us to evaluate observables
in the Einstein frame obtained after eliminating nonminimal couplings. 
Similarly, 
the tensor metric perturbations,
$h_{ij}$,
are also manifestly conformally invariant.

Similarly, 
the disformal transformation \cite{Bekenstein:1992pj,Zumalacarregui:2013pma,Bettoni:2013diz}
\be
\label{general_first}
{\tilde g}_{\mu\nu}
=\C(\phi,\X)g_{\mu\nu}+\D(\phi,\X) \phi_\mu \phi_\nu,
\ee
where 
$\C$ and $\D$ are the free functions of
the scalar field $\phi$ and the kinetic term $\X$
(see Eq. \eqref{defx}),
is known as the most general frame transformation
which is composed of the scalar field and its first-order derivatives.
The transformation \eqref{general_first} 
maps 
a class of the Class-${}^2$N-I and Class-${}^3$N-I DHOST theories
to another class \cite{Achour:2016rkg,BenAchour:2016fzp,Langlois:2018dxi}.
The Horndeski theories \cite{Horndeski:1974wa,Deffayet:2011gz,ginf}
are framed 
by the subclass of the disformal transformation\eqref{general_first},
$\C=\C(\phi)$ and $\D=\D(\phi)$ \cite{Bettoni:2013diz},
and
the beyond-Horndeski theories
are done
by the subclass of the disformal transformation \eqref{general_first}
$\C=\C(\phi)$ and $\D=\D(\phi,\X)$
\cite{Gleyzes:2013ooa,Gleyzes:2014dya,Gleyzes:2014qga},
respectively.

The invariance of the comoving curvature perturbation
within the above class of the disformal transformation
has been shown
in Refs.~\cite{Minamitsuji:2014waa,Tsujikawa:2014uza,Motohashi:2015pra,Domenech:2015hka}.
The tensor perturbations are always shown to be disformally invariant.
On the other hand, 
the invariance of the comoving curvature perturbation
depends on the subclass of the disformal transformation.
In the class of $\C=\C(\phi)$ and $\D=\D(\phi,\X)$,
the disformal invariance of the comoving curvature perturbation always holds
\cite{Minamitsuji:2014waa,Tsujikawa:2014uza}.
In the most general class that $\C=\C(\phi,\X)$ and $\D=\D(\phi,\X)$,
the invariance of the comoving curvature perturbation
holds approximately
on the superhorizon scales $k/(aH)\ll  1$ \cite{Motohashi:2015pra,Domenech:2015hka},
where $H(t):=\dot{a}/a$ represents the Hubble expansion rate,
whenever the gauge-invariant perturbation about the scalar field
\begin{eqnarray}
\label{sigma}
\Sigma
&:=& 
-\dot{\phi}_0
  \ddot{\phi}_0
\left(
\frac{\delta \X}{\dot{\X}_0}
-\frac{\phi_1}{\dot{\phi}_0}
\right)
=
  A\dot{\phi}_0^2
+ \phi_1\ddot{\phi}_0
-\dot{\phi}_0\dot{\phi}_1,
\end{eqnarray}
is suppressed on the superhorizon scales $k/(aH)\ll 1$,
where
$\X_0$ and $\delta \X$
represent the background and perturbation parts of $\X$
given, respectively,
by 
\begin{eqnarray}
\label{decomp2}
\X_0
:= -\dot{\phi}_0^2,
\qquad 
\delta \X
:=
2\dot{\phi}_0
\left(
A\dot{\phi}_0
-\dot{\phi}_1
\right).
\end{eqnarray}
Following the definition in Eq. \eqref{sigma},
$\Sigma$ represents the relative perturbation
between the scalar field and its kinetic term.
In the inflation models 
with the canonical/ noncanonical kinetic terms,
$\Sigma$ is proportional to the intrinsic entropy perturbation
of the scalar field
\be
\delta \Gamma_{(\phi)}
=\delta p_{(\phi)}
-\left(\frac{\dot{p}_{0(\phi)}(t)}{\dot{\rho}_{0(\phi)}(t)}\right)
 \delta \rho_{(\phi)},
\ee
where
$\left(\rho_{0(\phi)} (t), p_{0(\phi)} (t)\right)$
and 
$\left(\delta \rho_{(\phi)}, \delta p_{(\phi)}\right)$
represent the background and perturbation parts of
the energy density and pressure of the scalar field, respectively
\cite{Gordon:2000hv,Christopherson:2008ry}.
For this reason,
we call $\Sigma$
the intrinsic entropy perturbation of the scalar field.
In the models
in the Horndeski theories
with the canonical or noncanonical kinetic terms,
it has been shown that 
when the comoving curvature perturbation
is conserved on the superhorizon scales, 
\be 
\dot{\R}_c\approx 0,
\ee
where `$\approx$' means that
the equality holds only on the superhorizon scales,
the perturbation $\Sigma$ is suppressed on the superhorizon scales,
\be
\Sigma\approx 0,
\ee
(see Refs. \cite{Gordon:2000hv,Christopherson:2008ry,Minamitsuji:2014waa,Motohashi:2015pra}).
The results in Ref. \cite{Motohashi:2015pra}
suggest 
that also in the Class-${}^2$N-I and Class-${}^3$N-I DHOST theories,
when $\Sigma$ is suppressed on the superhorizon scales,
which are related to the Horndeski theory,
$\R_c$ is conserved on the superhorizon scales
via the disformal transformation \eqref{general_first}
\cite{Achour:2016rkg,BenAchour:2016fzp,Langlois:2018dxi}.
On the other hand, 
in the more general scalar-tensor theories with more than the third- order time derivatives,
the correspondence between the conservation of $\R_c$
and the suppression of $\Sigma$ on the superhorizon scales
has not been clarified yet.
Since the scalar-tensor theories with the third- order time derivatives
mentioned below have not been formulated yet,
for now we leave this subject for future work.

In this paper, 
we consider the generalized disformal transformation
with the second-order covariant derivatives of the scalar field.
First, we list the most fundamental scalar quantities
constructed with the covariant derivatives of the scalar field
with at most the quadratic order of  the second-order derivatives.
By ``fundamental'',
we mean
that all the other scalar quantities constructed with the derivatives of the scalar field
can be expressed as the products of them.
For instance, 
only the fundamental scalar quantity
constructed with the first-order derivatives
of the scalar field is 
given by the kinetic term $\X$, Eq. \eqref{defx}.
The other scalar quantities 
with the first-order derivatives
can be expressed in terms of the nonlinear combination of $\X$,
e.g., $\phi^\mu\phi^\nu(\phi_\mu\phi_\nu)=\X^2$.

At the linear order of the second-order covariant derivatives of the scalar field,
there are the two fundamental scalar quantities composed of the covariant derivatives of the scalar field
given by 
\begin{eqnarray}
\label{defbox}
\Box\phi
&:=&
 g^{\mu\nu} \times \phi_{\mu \nu},
\\
\label{defy}
\Y
&:= &
2
\phi^\mu
\phi^\nu
\times
\phi_{\mu\nu}
=
g^{\mu\nu}
\phi_\mu  \X_\nu,
\end{eqnarray}
where we have introduced the shorthand notation for
the covariant derivative of the kinetic term $\X_\mu:= \nabla_\mu \X$.
At the quadratic order of the second-order covariant derivatives of the scalar field,
there are also two
the fundamental scalar quantities composed of the covariant derivatives of the scalar field 
given by
\begin{eqnarray}
\label{defz}
\Z
&:=&
4g^{\mu\nu}
\phi^{\rho} 
\phi^{\sigma}
\times
\left(
\phi_{\mu\rho}
\phi_{\nu\sigma}
\right)
=
g^{\mu\nu}
\X_\mu 
\X_\nu,
\\
\label{defw}
\W
&:=&
g^{\mu\nu}
g^{\rho\sigma}
\times
\left(
\phi_{\mu\rho}
\phi_{\nu\sigma}
\right).
\end{eqnarray}
The other combinations of the covariant derivatives of the scalar field
with the quadratic order of the second-order covariant derivatives
can be expressed in terms of the fundamental quantities
at the linear order \eqref{defbox}-\eqref{defy},
e.g.,
$
\phi^{\rho}
\phi^\mu
\phi^{\sigma}
\phi^{\nu}
\times
\left(
\phi_{\mu\rho}
\phi_{\nu\sigma}
\right)
=\Y^2/4$
and 
$g^{\mu\rho}
\phi^\nu
\phi^\sigma
\times
\left(
\phi_{\mu\rho}
\phi_{\nu\sigma}
\right)
=\left(\Box\phi \right)\Y/2$.
We assume that the free functions
in the general disformal transformation 
are the functions of the fundamental elements
\eqref{defx}, and \eqref{defbox}-\eqref{defw},
as well as the scalar field itself $\phi$.

We also consider 
the tensors constructed with the covariant derivatives of the scalar field
\begin{eqnarray}
\label{disform_elements}
\phi_\mu \phi_\nu,
\quad
\phi_{\mu\nu},
\quad
\phi_{(\mu} \X_{\nu)},
\quad
\X_\mu \X_\nu,
\quad
\phi^{\rho}{}_{\mu} \phi_{\rho\nu},
\end{eqnarray}
whose contraction gives rise to the above fundamental scalar quantities 
$\X$, $\Box\phi$, $\Y$, $\Z$, and $\W$
(Eqs. \eqref{defx}, and \eqref{defbox}-\eqref{defw}),
respectively,
and assume
that Eqs. \eqref{disform_elements}
constitute the nonconformal part of the generalized disformal transformation.
By multiplying the free scalar functions of 
$\phi$, $\X$, $\Box\phi$, $\Y$, $\Z$, and $\W$
to the conformal and nonconformal parts,
we arrive at the disformal transformation considered in this paper
\begin{eqnarray}
\label{disform}
{\tilde g}_{\mu\nu}
&:=& 
\F_0 g_{\mu\nu}
+
\F_1
\phi_\mu \phi_\nu
+ 
\F_2
\phi_{\mu\nu}
+
\F_3
\phi_{(\mu} \X_{\nu)}
\nonumber\\
&+&
\F_4
\X_\mu \X_\nu
+
\F_5
g^{\rho\sigma}
\phi_{\rho\mu} \phi_{\sigma\nu},
\end{eqnarray}
where
$\F_I
:=
\F_I[
\phi,
\X,
\Box\phi,
\Y,
\Z,
\W]$
($I=0,1,2,3,4,5$).
Eq. \eqref{disform}
manifestly includes all the classes of the disformal transformation
constructed with the covariant derivatives of the scalar field
with at most the first-order covariant derivatives \eqref{general_first}.
On the other hand, 
Ref. \cite{Alinea:2020laa} considered 
the disformal transformation 
with the second-order covariant derivatives of the scalar field
given by 
\begin{eqnarray}
\label{disform0}
{\tilde g}_{\mu\nu}
=
\G_0 g_{\mu\nu}
+
\left(
\G_1 \phi_\mu
+
\G_2 X_\mu
\right)
\left(
\G_1 \phi_\nu
+
\G_2 X_\nu
\right),
\end{eqnarray}
where
$\G_J
:=\G_J
[\phi,
\X,
\Y,
\Z]$
($J=0,1,2,3$).
The correspondence between Eqs. \eqref{disform} and \eqref{disform0}
is given by 
$\F_0=\G_0$,
$\F_1=\G_1^2$,
$\F_2=0$,
$\F_3=2\G_1\G_2$,
$\F_4=\G_2^2$,
and 
$\F_5=0$.
Thus, the transformation \eqref{disform} generalizes Eq. \eqref{disform0},
in terms of 
the additional dependence on $\Box\phi$ and $\W$,
and 
the additional nonconformal parts
$\phi_{\mu\nu}$
and
$\phi^{\rho}{}_{\mu} \phi_{\rho\nu}$ in Eq. \eqref{disform_elements}.
As we will see in Sec. \ref{sec4},
these new terms result in 
the difference in the tensor perturbations
between the frames,
when the tensor perturbations are not conserved.

A crucial difference of the generalized disformal transformation with
the second-order covariant derivatives of the scalar field
from the disformal transformation only with the first-order derivatives \eqref{general_first}
is 
that there will be infinite number of terms 
that constitute the transformation.
For instance, 
we may add
the more than the cubic order powers 
of the second-order covariant derivative of the scalar field,
e.g.,  
$\phi_{\mu}{}^{\rho}\phi_{\rho}{}^{\alpha}\phi_{\alpha\nu}$,
$\phi_{\mu}{}^{\rho}\phi_{\rho}{}^{\alpha}\phi_{\alpha}{}^{\beta}\phi_{\beta\nu}$,
and so on in Eq. \eqref{disform}.
In order to scan the full space of the disformal transformation
with the second-order covariant derivatives of the scalar field,
a step-by-step analysis
by adding a higher-order power of 
the second-order covariant derivatives of the scalar field
would be necessary.
On the other hand, 
higher-order derivative couplings 
to the matter and gravity sectors in the cosmological backgrounds
would lead to anomalous behaviours,
e.g., the partial breaking of the screening mechanism inside the stars
~\cite{Kobayashi:2014ida,Koyama:2015oma,Saito:2015fza}
and
the decay of gravitational waves
\cite{Creminelli:2018xsv,Creminelli:2019kjy},
which would severely constrain the theory from the observational viewpoints,
although these constraints may not be applied to the models of
inflation and early Universe. 
Thus, 
to what extent we should extend the framework of the disformal transformation 
is indeed the matter of interests.
We would like to emphasize that 
the framework of the generalized disformal transformation \eqref{disform} is self-contained,
and 
sufficient to see the essential features of the disformal transformation 
with the second-order covariant derivatives of the scalar field.

After the generalized disformal transformation \eqref{disform},
the action of the scalar-tensor theory
written
in terms of the new frame metric ${\tilde g}_{\mu\nu}$
and the scalar field $\phi$
would contain the third-order covariant derivatives
of the scalar field ${\tilde \nabla}_\mu {\tilde \nabla}_\nu {\tilde \nabla}_\alpha\phi$,
where ${\tilde \nabla}_\mu$ denotes the covariant derivative
associated with the new metric ${\tilde g}_{\mu\nu}$,
and
after the Arnowitt-Deser-Misner  (ADM) decomposition \cite{Arnowitt:1959ah},
the third-order time derivative terms such as ${\cal H} (\phi,\dot{\phi},\ddot{\phi})
\dddot{\phi}^2$,
which would lead to the equations of motion with the sixth-order time derivatives,
and hence the two Ostrogradsky ghosts.
To our knowledge,
the degenerate scalar-tensor theories
with the third- and higher order covariant derivative terms of the scalar field
have not been constructed yet.
Although the construction of these scalar-tensor theories is not our main purpose,
we would like to mention the properties of 
the theory with the third-order time derivatives within analytical mechanics.

In Appendix \ref{app_a},
we show
a simple example of the degenerate theory
with the third-order time derivative in analytical mechanics. 
Refs. \cite{Motohashi:2017eya,Motohashi:2018pxg}
discussed
more general properties of
analytical mechanics with the third- and higher-order time derivatives,
and
obtained the conditions to avoid the Ostrogradsky ghosts.
In analytical mechanics with the second-order time derivatives, 
eliminating the linear momentum terms in the Hamiltonian
by the secondary constraints 
that ensure the time evolution of the primary constraints arising from the degeneracy conditions  
is enough
to remove all the the Ostrogradsky ghosts
\cite{Motohashi:2014opa,Langlois:2015cwa,Motohashi:2016ftl,Klein:2016aiq}.
On the other hand, 
in analytical mechanics
with more than the third-order time derivatives, 
eliminating all the linear momentum terms from the Hamiltonian
is not enough
and 
more secondary constraints
are necessary
to remove all the Ostrogradsky ghosts.
The extension of analytical mechanics  
with more than the third-order time derivatives
to the scalar-tensor theories 
would also be the nontrivial issue 
from both the theoretical and technical aspects.
In this paper, 
we will not focus on the construction of 
the scalar-tensor theories 
with more than the third-order time derivative terms
without the Ostrogradsky ghosts,
and 
simply assume 
that these scalar-tensor theories exist
and 
the subclass of them
is related 
via the generalized disformal transformation \eqref{disform}
as in the case of analytical mechanics.

In Sec. \ref{sec2},
we will derive the generalized disformal transformation of the scalar perturbations. 
In Sec. \ref{sec3},
we will derive the generalized disformal transformation \eqref{disform}
of the comoving curvature perturbation
and 
the conditions 
under which 
the comoving curvature perturbations are disformally invariant
on the superhorizon scales.
In Sec. \ref{sec4},
we discuss the generalized disformal transformation \eqref{disform}
of the tensor perturbations.
The last Sec. \ref{sec5}
will be devoted to giving a brief summary and conclusion.

\section{The scalar perturbations}
\label{sec2}

First, we consider the disformal transformation of scalar perturbations.
Following the decomposition \eqref{scalar_flrw} and \eqref{decomp1},
the fundamental scalar quantities in the generalized disformal transformation \eqref{disform}
are decomposed 
into the background and perturbation parts as
Eq. \eqref{decomp2}, and 
\begin{eqnarray}
\label{decomp3}
\Box\phi
&=&
-
\left(
\ddot{\phi}_0
+3\frac{\dot{a}}{a}
\dot{\phi}_0
\right)
\nonumber\\
&+&
2\left(
 \ddot{\phi}_0
+3\frac{\dot{a} \dot{\phi}_0}{a}
\right) 
A
+
\left(
-\ddot{\phi}_1
-3\frac{\dot{a}}{a} \dot{\phi}_1
\right)
\nonumber\\
&+&
\dot{\phi}_0
 \dot{A}
+3\dot{\phi}_0
 \dot{\psi}
-k^2
\frac{1}{a^2}
(\delta_g\phi),
\end{eqnarray}
\begin{eqnarray}
\label{decomp4}
\Y
&=&
2\dot{\phi}_0^2\ddot{\phi}_0
\nonumber\\
&+&
2
\dot{\phi}_0
\left(
-4A\dot{\phi}_0 \ddot{\phi}_0
-\dot{\phi}_0^2 \dot{A}
+2\ddot{\phi}_0\dot{\phi}_1
+\dot{\phi}_0 \ddot{\phi}_1
\right),
\end{eqnarray}
\begin{eqnarray}
\label{decomp5}
\Z
&=&
-4\dot{\phi}_0^2\ddot{\phi}_0^2
\nonumber\\
&+&
8\dot{\phi}_0\ddot{\phi}_0
\left(
3A \dot{\phi}_0\ddot{\phi}_0
+\dot{\phi}_0^2 \dot{A}
-\ddot{\phi}_0
  \dot{\phi}_1
-\dot{\phi}_0
  \ddot{\phi}_1
\right),
\end{eqnarray}
\begin{eqnarray}
\label{decomp6}
\W
&=&
\left(
\ddot{\phi}_0^2
+
3\frac{\dot{a}^2}{a^2}\dot{\phi}_0 ^2
\right)
\nonumber\\
&-&
4A
\left(
\ddot{\phi}_0^2
+3 \frac{\dot{a}^2}{a^2}\dot{\phi}_0^2
\right)
-
2\dot{\phi}_0\ddot{\phi}_0
 \dot{A}
+6
\frac{\dot{a}^2\dot{\phi}_0\dot{\phi}_1}
       {a^2}
\nonumber\\
&-&
6
\frac{\dot{a}\dot{\phi}_0^2\dot{\psi}}
       {a}
+2
\ddot{\phi}_0
\ddot{\phi}_1
+
2k^2\frac{\dot{a} \dot{\phi}_0}{a^3}
(\delta_g\phi),
\end{eqnarray}
where we have introduced 
the gauge-invariant scalar field perturbation in the longitudinal gauge
($B=E=0$),
\be
\delta_g\phi
:=
 \phi_1
-a^2  \dot{\phi}_0
\left(
\dot{E}
-\frac{B}{a}
\right).
\ee
Accordingly,
the functions in Eq. \eqref{disform}
can be written as
\be
\F_I= F_I (t) +\delta \F_I(t,x^i),
\ee
where 
$
F_I (t)
=\F_I 
[\phi_0,
\X_0,
\Box_0\phi_0, 
\Y_0,
\Z_0,
\W_0]$
with the index ``0'' representing the background part of
Eqs. \eqref{decomp1}, \eqref{decomp2}, \eqref{decomp3}-\eqref{decomp6},
and 
\begin{eqnarray}
\delta\F_I
&:=&
F_{I,\phi}\delta\phi
+  
F_{I,\X}\delta \X
+  
F_{I,\Box\phi}\delta(\Box\phi)
\nonumber\\
&+&
F_{I,\Y}\delta \Y
+
F_{I,\Z}\delta\Z
+
F_{I,\W}\delta \W,
\end{eqnarray}
with 
$\delta\phi$,
$\delta \X$,
$\delta (\Box\phi)$,
$\delta \Y$,
$\delta \Z$,
and 
$\delta \W$
being the perturbation parts of
Eqs. \eqref{decomp1}, \eqref{decomp2}, and \eqref{decomp3}-\eqref{decomp6},
and 
$F_{I,\phi}:= \partial_\phi \F_I$,
$F_{I,\X}:= \partial_\X \F_I$,
$F_{I,\Box \phi}:= \partial_{\Box\phi} \F_I$,
$F_{I,\Y}:= \partial_\Y \F_I$,
$F_{I,\Z}:= \partial_\Z \F_I$,
and 
$F_{I,\W}:= \partial_\W \F_I$
evaluated at the background $[\phi_0,\X_0,\Box_0\phi_0, \Y_0,\Z_0,\W_0]$.

Under the generalized disformal transformation \eqref{disform},
the background part of Eq. \eqref{scalar_flrw}
is mapped to the in the form of the FLRW metric  
\be
d{\tilde s}_0^2
={\tilde g}_{0,\mu\nu}
 dx^\mu
 dx^\nu
=-d{\tilde t}^2
+{\tilde a}(\tit)^2 \delta_{ij}dx^i dx^j,
\ee
where we have defined
\be
d{\tilde t}:=\sqrt{\A(t)} dt,
\qquad
{\tilde a} ({\tilde t}):=\sqrt{\B(t) } a(t)
\ee
with the requirements $\A(t)>0$ and $\B(t)>0$,
respectively,
with 
\begin{eqnarray}
\A(t)
&:=&
F_0 
-
F_1 
\dot{\phi}_0^2
\nonumber\\
&-&
\ddot{\phi}_0
\left\{
F_2
-2F_3 \dot{\phi}_0^2
+
\left(
  4F_4 \dot{\phi}_0^2
-F_5
\right)
\ddot{\phi}_0
\right\},
\\ 
\B(t)
&:=&
 F_0
-\frac{\dot{a}}{a}\dot{\phi}_0 F_2
+ \frac{\dot{a}^2}{a^2}\dot{\phi}_0^2 F_5,
\end{eqnarray}
and the perturbed part of Eq. \eqref{scalar_flrw}
is disformally transformed as 
\be
\delta{\tilde{g}}_{\mu\nu}
dx^\mu dx^\nu
=
\delta {\tilde g}_{tt} dt^2
+2\delta {\tilde g}_{t i} dt dx^i
+\delta {\tilde g}_{ij} dx^i dx^j,
\ee
where
\begin{eqnarray}
\delta {\tilde g}_{tt}
&:=&
- \delta \F_0
+ \dot{\phi}_0^2 \delta\F_1
+\ddot{\phi}_0 \delta \F_2
-2\dot{\phi}_0^2\ddot{\phi}_0 \delta \F_3
\nonumber\\
&+&
4 \dot{\phi}_0^2 \ddot{\phi}_0^2 \delta \F_4
-\ddot{\phi}_0^2\delta \F_5
\nonumber\\
&-&
  2A F_0
+2\dot{\phi}_0 \dot{\phi}_1 F_1
+\left(-\dot{\phi}_0 \dot{A}_2 +\ddot{\phi}_1\right)
  F_2
\nonumber\\
&+&
2\dot{\phi}_0
\left(
2\dot{\phi}_0\ddot{\phi}_0 A
+\dot{\phi}_0^2\dot{A}
-2\ddot{\phi}_0 \dot{\phi}_1 
-\dot{\phi}_0\ddot{\phi}_1
\right)
F_3
\nonumber\\
&-&
8\dot{\phi}_0\ddot{\phi}_0
\left(
 2A\dot{\phi}_0\ddot{\phi}_0
+\dot{\phi}_0^2\dot{A}
-\ddot{\phi}_0 \dot{\phi}_1
-\dot{\phi}_0\ddot{\phi}_1
\right)
F_4
\nonumber\\
&+&
2\ddot{\phi}_0
\left(
\ddot{\phi}_0 A
+
\dot{\phi}_0
\dot{A}
-
\ddot{\phi}_1
\right)
F_5,
\end{eqnarray}
\begin{eqnarray}
\delta 
{\tilde g}_{ti}
&:=&
a(t)
\partial_i
\left[
   B F_0
+\frac{\dot{\phi}_0}{a} \phi_1 F_1
\right.
\nonumber\\
&&
\left.
+
\left\{
-\dot{\phi}_0 
 \left(
 \frac{A}{a}
+\frac{\dot{a}}{a} B
 \right)
-\frac{\dot{a}}{a^2} \phi_1
+\frac{\dot{\phi}_1}{a}
\right\}
F_2
\right.
\nonumber\\
&&
\left.
+
\frac{\dot{\phi}_0}{a}
\left(
 \dot{\phi}_0^2 A
-\ddot{\phi}_0 \phi_1
-\dot{\phi}_0 \dot{\phi}_1
\right)
F_3
\right.
\nonumber\\
&&
\left.
-4
\frac{\dot{\phi}_0^2
 \ddot{\phi}_0}
{a}
\left(
  \dot{\phi}_0 A
-\dot{\phi}_1
\right)
F_4
\right.
\nonumber\\
&&
\left.
+
\frac{1}{a}
\left(
 \frac{\dot{a}\dot{\phi}_0^2}{a}A
+\dot{\phi}_0\ddot{\phi}_0 A
+\frac{\dot{a}^2\dot{\phi}_0^2}{a} B
+\frac{\dot{a}^2\dot{\phi}_0}{a^2} \phi_1
\right.
\right.
\nonumber\\
&&
\left.
\left.
+\frac{\dot{a} \ddot{\phi}_0}
          {a}
 \phi_1
-\frac{\dot{a}\dot{\phi}_0}
          {a}
\dot{\phi}_1
-\ddot{\phi}_0
  \dot{\phi}_1
\right)
F_5
\right],
\end{eqnarray}
and 
\begin{eqnarray}
\delta\tilde{g}_{ij}
&:=&
a(t)^2
\left[
{\delta \F}_0
-\frac{\dot{a} \dot{\phi}_0}{a} 
{\delta \F}_2
+
\frac{\dot{a}^2\dot{\phi}_0^2}{a^2}
{\delta \F_5}
\right.
\nonumber\\
&-&
2\psi F_0
+
\left(
  2\frac{\dot{a}\dot{\phi}_0}{a}A
+2\frac{ \dot{a} \dot{\phi}_0}{a}\psi
-\frac{\dot{a}}{a}\dot{\phi}_1
+\dot{\phi}_0\dot{\psi}
\right)
F_2
\
\nonumber\\
&-&
\left.
\left(
4\frac{\dot{a}^2 \dot{\phi}_0^2}{a^2}  A
+2\frac{\dot{a}^2 \dot{\phi}_0^2}{a^2}  \psi
-2\frac{\dot{a}^2 \dot{\phi}_0}{a^2}\dot{\phi}_1
+2\frac{\dot{a}\dot{\phi}_0^2} {a} \dot{\psi}
\right)
F_5
\right]
\delta_{ij}
\nonumber\\
&+&
a(t)^2
\partial_i
\partial_j
\left[
2 E F_0
\right.
\nonumber\\
&+&
\left.
\left(
  \frac{\dot{\phi}_0}{a} B
-2\frac{\dot{a} \dot{\phi}_0}{a} E
+\frac{\phi_1}{a^2}
-\dot{\phi}_0 \dot{E}
\right)
F_2
\right.
\nonumber\\
&+&
\left.
\left(
-2\frac{\dot{a} \dot{\phi}_0^2 }{a^2} B
+2\frac{\dot{a}^2 \dot{\phi}_0^2}{a^2} E
-2\frac{\dot{a} \dot{\phi}_0}{a^3}
   \phi_1
+2\frac{\dot{a} \dot{\phi}_0^2}{a}
 \dot{E}
\right)
F_5
\right].
\nonumber\\
&&
\end{eqnarray}
By changing the time coordinate from $t$ to ${\tilde t}$,
$\delta {\tilde g}_{\tit \tit}= \frac{1}{\A} \delta {\tilde g}_{tt}$
and 
$\delta {\tilde g}_{\tit i }= \frac{1}{\sqrt{\A}} \delta {\tilde g}_{ti}$,
we define the metric perturbations after the generalized disformal transformation \eqref{disform}
in the same manner as Eq. \eqref{scalar_flrw}
by attaching `tilde' to all the perturbation variables in the new frame.
Thus, the metric perturbations in the new frame are given by 
\begin{eqnarray}
{\tilde A}
&:=&
\frac{1}{\A}
\left[
\frac{1}{2} \delta \F_0
- \frac{1}{2}\dot{\phi}_0^2 \delta\F_1
-\frac{1}{2}\ddot{\phi}_0 \delta \F_2
+\dot{\phi}_0^2\ddot{\phi}_0 \delta \F_3
\right.
\nonumber\\
&-&
2\dot{\phi}_0^2 \ddot{\phi}_0^2 \delta \F_4
+\frac{1}{2}\ddot{\phi}_0^2\delta \F_5
\nonumber\\
&+&
  A F_0
-\dot{\phi}_0 \dot{\phi}_1 F_1
-\frac{1}{2}
\left(-\dot{\phi}_0 \dot{A}_2 
+\ddot{\phi}_1\right)
  F_2
\nonumber\\
&-&
\dot{\phi}_0
\left(
2\dot{\phi}_0\ddot{\phi}_0 A
+\dot{\phi}_0^2\dot{A}
-2\ddot{\phi}_0 \dot{\phi}_1 
-\dot{\phi}_0\ddot{\phi}_1
\right)
F_3
\nonumber\\
&+&
4\dot{\phi}_0\ddot{\phi}_0
\left(
 2A\dot{\phi}_0\ddot{\phi}_0
+\dot{\phi}_0^2\dot{A}
-\ddot{\phi}_0 \dot{\phi}_1
-\dot{\phi}_0\ddot{\phi}_1
\right)
F_4
\nonumber\\
&-&
\left.
\ddot{\phi}_0
\left(
\ddot{\phi}_0 A
+
\dot{\phi}_0
\dot{A}
-
\ddot{\phi}_1
\right)
F_5
\right],
\end{eqnarray}
and 
\begin{eqnarray}
{\tilde \psi}
&=&
\frac{1}{\B}
\left[
-\frac{1}{2}
{\delta \F}_0
+
\frac{\dot{a} \dot{\phi}_0 }{2a} 
{\delta \F}_2
-
\frac{\dot{a}^2 \dot{\phi}_0^2 }{2a^2}
{\delta \F_5}
\right.
\nonumber\\
&&
\left.
+\psi F_0
-
\left(
  \frac{\dot{a}}{a}\dot{\phi}_0 A
+\frac{ \dot{a} }{a}\dot{\phi}_0\psi
-\frac{\dot{a}}{2a}\dot{\phi}_1
+\frac{1}{2}\dot{\phi}_0\dot{\psi}
\right)
F_2
\right.
\nonumber\\
&&
\left.
-
\left(
-2\frac{\dot{a}^2  \dot{\phi}_0^2}{a^2} A
-\frac{\dot{a}^2 \dot{\phi}_0^2 }{a^2} \psi
+\frac{\dot{a}^2 \dot{\phi}_0}{a^2} \dot{\phi}_1
-\frac{\dot{a} \dot{\phi}_0^2} {a} \dot{\psi}
\right)
F_5
\right].
\nonumber\\
\end{eqnarray}

\vspace{0.3cm}

\section{The comoving curvature perturbations}
\label{sec3}

We define the comoving curvature perturbation in the new frame,
given by 
\begin{eqnarray}
\label{com2}
{\tilde \R}_c
=\psi 
+\frac{1}
{\phi_{0,\tit}}
\frac{\tia_{,\tit}}{\tia}
\phi_1,
\end{eqnarray}
which is shown to be gauge-invariant.
A straightforward computation shows that
the difference between the comoving curvature perturbations 
in the original and new frames,
Eqs. \eqref{com1} and \eqref{com2},
is given by the combination of the gauge-invariant perturbations
in the original frame
\begin{eqnarray}
\label{diff1}
&&
\tilde{\R}_c
-\R_c
=
\frac{1}
       {a^2F_0- a\dot{a}\dot{\phi}_0 F_2+ \dot{a}^2 \dot{\phi}_0^2F_5}
\times
\nonumber\\
&&
\left(
\Q_1 (t)
\dot{\R}_c
+
\Q_2(t)
\Sigma
+
\Q_3 (t)
\dot{\Sigma}
-
k^2
\Q_4(t)
\left(\delta_g\phi\right)
\right),
\end{eqnarray}
where
the background-dependent functions 
$\Q_1(t)$, $\Q_2(t)$, $\Q_3(t)$, and $\Q_4(t)$
are,
respectively, 
given 
in Eqs. \eqref{q1}-\eqref{q4}
in Appendix \ref{app_b}.
Thus,  
in the case that 
\begin{itemize}
\label{condcond}

\item{
(1) the adiabaticity holds on the superhorizon scales, 
$\Sigma\approx 0$ and $\dot{\Sigma}\approx 0$,}

\item 
{(2)
the comoving curvature perturbation in the original frame $\R_c$ is conserved
on the superhorizon scales, $\dot{\R}_c\approx 0$,}

\end{itemize}
we 
find that the comoving curvature perturbations in both the frames 
coincide on the superhorizon scales $k/(aH)\ll  1$,
\be
{\tilde \R}_{c} \approx  \R_{c},
\ee
and the equivalence of the comoving curvature perturbations
on the superhorizon scales.
This is a direct extension of the results obtained in Ref. \cite{Alinea:2020laa}.

As mentioned in Sec. \ref{sec1}, 
in the various single-field inflation models,
the conditions (1) and (2) 
hold at the same time  on the superhorizon scales
\cite{Gordon:2000hv,Christopherson:2008ry,Minamitsuji:2014waa,Motohashi:2015pra}.
In these models,
when $\Sigma\approx 0$ and $\dot{\Sigma}\approx 0$, 
the decaying mode among the two independent solutions 
of ${\cal R}_c$
is negligible on the superhorizons scales
and $\dot{\R}_c\approx 0$.
On the other hand,
in the more general scalar-tensor models,
e.g., in the DHOST theories and 
the theories with more than the third--order time derivatives,
the correspondence between the conditions (1) and (2) 
has not been clarified yet.
The exceptional case is 
that the background-dependent coefficients
vanish coincidently,
$\Q_1(t)=0$, $\Q_2(t)=0$, $\Q_3(t)=0$, and $\Q_4(t)=0$,
where the exact frame invariance $\tilde{\R}_c=\R_c$ holds 
at any scale,
irrespective of the adiabaticity of the scalar field.
Since there are the six free functions $\F_I$ ($I= 0,1,2,3,4,5$),
it is able to achieve this condition.

It is also straightforward to confirm that
the gauge-invariant
intrinsic entropy perturbation in the new frame,
denoted by ${\tilde \Sigma}$,
can be written 
in terms of the gauge-invariant perturbations in the original frame 
\begin{eqnarray}
\label{diff2}
{\tilde \Sigma}
&:=&
{\tilde A}
\phi_{0,{\hat t}}^2
+\phi_1
\phi_{0,{\hat t}{\hat t}}
-\phi_{1,{\hat t}}
 \phi_{0,{\hat t}}
\nonumber
\\
&=&
\frac{
\alpha_1 (t)
\Sigma
+
\alpha_2(t)
\dot{\Sigma}
+
\alpha_3 (t)
\dot{\R}_c
-
k^2
\alpha_4(t)
\left(\delta_g\phi \right)}
{{\A}^2},
\end{eqnarray}
where $\alpha_1(t)$, $\alpha_2 (t)$, $\alpha_3 (t)$, and $\alpha_4 (t) $
are the background-dependent coefficients
given
in Eqs. \eqref{a1}-\eqref{a4} in Appendix \ref{app_b},
respectively.
Thus, 
whenever conditions (1) and (2) hold in the original frame,
$\dot{\R}_c\approx 0$,
the adiabaticity also holds
in the new frame ${\tilde \Sigma}\approx 0$.

Now, we confirm that the previous results 
for the disformal transformation 
with the first-order derivative of the scalar field
\eqref{general_first}
can be reproduced.
\begin{itemize}

\item
In the case
\be
\F_0=\F_0(\phi), 
\qquad
\F_1=\F_1(\phi,\X),
\qquad
\F_I=0,
\ee
 ($I=2,3,4,5$)
with $\A=F_0-F_1\dot{\phi}_0^2$ and $\B=F_0$,
which 
relates a class of the Horndeski \cite{Horndeski:1974wa,Deffayet:2011gz,ginf} 
and beyond-Horndeski \cite{Gleyzes:2013ooa,Gleyzes:2014dya,Gleyzes:2014qga} theories
to another \cite{Bettoni:2013diz}.
we obtain $\Q_1(t)=\Q_2(t)=\Q_3(t)=\Q_4(t)=0$
and hence the relation 
\be
{\tilde \R}_{c} =\R_{c},
\ee
which shows 
that the disformal invariance
of the comoving curvature perturbations 
exactly holds at any scale \cite{Minamitsuji:2014waa,Tsujikawa:2014uza}.

\item
In the case that 
\be
\F_0=\F_0(\phi,\X), 
\qquad
\F_1=\F_1(\phi,\X),
\qquad
\F_I=0,
\ee
($I=2,3,4,5$)
with $\A= F_0-F_1\dot{\phi}_0^2$ and $\B=F_0$,
which relates a class of the Class-${}^2$N-I and Class-${}^3$N-I DHOST theories
to another \cite{Achour:2016rkg,BenAchour:2016fzp,Langlois:2018dxi},
we find that $\Q_1=\Q_3=\Q_4=0$
and $\Q_2=-a^2 F_{0\X}$,
and hence the relation
\be
{\tilde \R}_{c}
-\R_{c}
=-\frac{F_{0X}} 
           {F_0}
\Sigma,
\ee
which confirms the results in Refs. \cite{Motohashi:2015pra,Domenech:2015hka}.
Thus,
in the Class-${}^2$N-I and Class-${}^3$N-I DHOST theories,
the equivalence between ${\cal R}_c$ and ${\tilde {\cal R}}_c$ holds 
when the intrinsic entropy perturbation $\Sigma$
is suppressed on the superhorizon scales.

\end{itemize}

Since 
$\R_c$, 
$\Sigma/\left(\dot{\phi_0}\right)^2$, 
and 
$\delta_g \phi/(-a^2 \dot{\phi}_0)$
are
all the gauge-invariant versions
of the metric perturbations 
$\psi$,
$A$,
and
$\dot{E}-B/a$
constructed
as the combinations
with the scalar field perturbation $\phi_1$,
we expect that 
the form of the relations between frames 
\eqref{diff1} and \eqref{diff2}
would remain the same 
even 
for the more general disformal transformations than Eq. \eqref{disform},
although
the background-dependent coefficients 
$\Q_1(t)$, $\Q_2(t)$, $Q_3(t)$, $\Q_4(t)$,
$\alpha_1(t)$, $\alpha_2(t)$, $\alpha_3(t)$, and $\alpha_4(t)$
are modified accordingly.
Thus, 
whenever 
in the original frame
on the superhorizon scales
the conditions (1) and (2) are satisfied at the same time
we expect that 
the comoving curvature perturbation
is disformally invarint on the superhorizon scales,
$\tilde{\R}_c\approx \R_c$
even for the more general disformal transformations.

\section{The tensor perturbations}
\label{sec4}

We then consider the tensor perturbations about 
a FLRW spacetime \eqref{scalar_flrw}.
The functions 
$\X$,
$\Box\phi$,
$\Y$,
$\Z$,
$\W$,
and hence 
$\F_I$ ($I=0,1,2,3,4,5$)
in the generalized disformal transformation \eqref{disform}
remain the same as the background
against the tensor perturbations.

Defining the tensor perturbations in the new frame ${\tilde h}_{ij}$
as in the same manner of the scalar perturbations,
$\delta {\tilde g}_{ij} dx^i dx^j
={\tilde a}^2({\tilde t})\tilde{h}_{ij}({\tilde t},x^i) dx^i dx^j$,
the difference between $h_{ij}$ and ${\tilde h}_{ij}$ 
can be computed as 
\be
\label{diff3}
{\tilde h}_{ij}
-h_{ij}
=
-
\frac{a\dot{\phi}_0 \left(aF_2-2\dot{a}\dot{\phi}_0F_5\right)}
{2\left(a^2F_0-a\dot{a} \dot{\phi}_0 F_2+\dot{a}^2\dot{\phi}_0^2F_5\right)}
\dot{h}_{ij},
\ee
where the nonzero difference 
arises from the disformal elements
$\phi_{\mu\nu}$
and
$g^{\rho\sigma}\phi_{\rho\mu} \phi_{\sigma\nu}$
in Eq, \eqref{disform_elements},
which were not discussed in Ref. \cite{Alinea:2020laa}.
This is because for instance 
the $\phi_{ij}$ term in the spatial components of Eq. \eqref{disform}
gives rise to the nonzero contribution 
to the difference as $\sim \dot{h}_{ij}\dot{\phi}_0$.
Eq.\eqref{diff3}
also confirms that 
the tensor pertubations in the new frame
also obeys 
the transverse-traceless gauge conditions 
$\delta^{ij} {\tilde h}_{ij}=\partial^i {\tilde h}_{ij}=0$.

Thus,
the tensor perturbations 
are also disformally invariant,
in the case that 
the tensor perturbations in the original frame $h_{ij}$
are conserved with time,
$\dot{h}_{ij}=0$.
The exceptional case is 
that the background-dependent coefficient
$aF_2-2\dot{a}\dot{\phi}_0F_5(t)=0$
or $\dot{\phi}_0=0$,
where the exact frame invariance $\tilde{h}_{ij}=h_{ij}$ is obtained
even if $\dot{h}_{ij}\neq 0$.

We expect 
that 
even for the more general disformal transformation
than Eq. \eqref{disform},
the difference in the tensor perturbations between frames 
is proportional to $\dot{h}_{ij}$.
This is because
at the level of the linearized perturbations
the nonzero difference between $h_{ij}$ and ${\tilde h}_{ij}$
would always 
arise
from the contributions as
$\delta\Gamma_{ij}^t\dot{\phi}_0$ and 
$\delta \Gamma_{it}^j\dot{\phi}_0$,
both of which are proportional to $\dot{h}_{ij}$.
Thus, 
even in the case of the more general disformal transformation
than Eq.~\eqref{disform},
whenever the tensor perturbations are 
conserved on the superhorizon scales, 
$\dot{h}_{ij}\approx 0$,
the tensor perturbations
remain invariant 
on the superhorizon scales,
${\tilde h}_{ij}\approx h_{ij}$.

\section{Conclusions}
\label{sec5}

We have investigated 
how the comoving curvature perturbation and tensor perturbations 
are transformed 
under the generalized disformal transformation 
with the second-order covariant derivatives of the scalar field.
To construct the generalized disformal transformation,
we considered
the fundamental elements \eqref{defx} and \eqref{defbox}-\eqref{defw}
constructed with the covariant derivatives of the scalar field
with at most the quadratic order of the second-order covariant derivatives 
of the scalar field, 
and the covariant tensors
whose contraction gives rise to the above fundamental elements.
The resultant general form of the disformal transformation
was given by Eq. \eqref{disform},
which included all the models
with the disformal transformation \eqref{general_first}
studied previously.

We then defined the gauge-invariant comoving curvature perturbations
both in the original and new frames
defined by Eqs. \eqref{com1} and \eqref{com2},
and computed their difference.
While reproducing the previous results on the disformal invariance
in Refs.~\cite{Minamitsuji:2014waa,Tsujikawa:2014uza,Motohashi:2015pra,Domenech:2015hka,Alinea:2020laa}, 
we have also shown that the difference between the comoving curvature perturbations
in the original and new frames,
${\cal R}_c$ and $\tilde{\cal R}_c$,
was given by the combination of the time derivative of the comoving curvature perturbation
in the original frame $\dot{\R}_c$, 
the gauge-invariant perturbation $\Sigma$ given by Eq. \eqref{sigma},
which is related to the intrinsic entropy perturbation of the scalar field,
and its time derivative.
In the case that 

\begin{itemize}
\label{condcond}

\item{
(1) the adiabaticity holds on the superhorizon scales, 
$\Sigma\approx 0$ and $\dot{\Sigma}\approx 0$,}

\item 
{(2)
the comoving curvature perturbation in the original frame $\R_c$ is conserved
on the superhorizon scales, $\dot{\R}_c\approx 0$,}

\end{itemize}
the equivalence of the comoving curvature perturbations
under the generalized disformal transformation \eqref{disform}
holds on the superhorizon scales.
While in the previously known scalar-tensor theories,
whenever the condition (1) holds the condition (2) also hods, 
in the more general scalar-tensor theories with the third--order time derivatives,
which are related via the generalized disformal transformation \eqref{disform} 
the relationship between the conditions (1) and (2) has not been clarified yet.
Thus, 
in this paper,
we raise the conditions (1) and (2) as the independent ones.

We have also shown
that the difference between the tensor perturbations
was proportional to the time derivative 
of the tensor perturbations in the original frame.
Thus,
the tensor perturbations 
were also disformally invariant,
whenever
the tensor perturbations in the original frame
were conserved with time.

We should emphasize again that 
the disformal transformation Eq. \eqref{disform} is not the most general one,
in terms of the power of the second-order covariant derivatives
and the order of the highest-order derivatives in the transformation.
It would be interesting and important to 
extend the analysis in this paper
to these more general disformal couplings.
We hope to come back to these issues in our future work.

\acknowledgments{
M.M.~was supported by the Portuguese national fund 
through the Funda\c{c}\~{a}o para a Ci\^encia e a Tecnologia
in the scope of the framework of the Decree-Law 57/2016 of August 29
(changed by Law 57/2017 of July 19),
and by the CENTRA through the Project~No.~UIDB/00099/2020.
}

\appendix

\section{A degenerate theory with third-order time derivatives in analytical mechanics}
\label{app_a}

We consider the theory with the third-order time derivative
in analytical mechanics
\be
\label{3rd1}
L
&=&
\frac{a_1}{2}\dddot{\phi}^2
+\frac{a_2}{2} \ddot{\phi}^2
+\frac{a_3}{2}\dot{\phi}^2
+\frac{b_1}{2}\dot{q}^2
+b_2 \dddot{\phi} \dot{q}
\nonumber\\
&-&V(\phi,q),
\ee
where
$a_1$, $a_2$, $a_3$, $b_1$, and $b_2$
are constants, 
and the potential is given by the quadratic terms
\begin{eqnarray}
V(\phi,q)
=\frac{g_1}{2}\phi^2
+g_2 \phi q
+\frac{g_3}{2}q^2,
\end{eqnarray}
with $g_1$, $g_2$, and $g_3$ being constants.
The theory equivalent to Eq. \eqref{3rd1}
can be obtained
by introducing the two auxiliary fields $R$ and $Q$ 
\be
\label{3rd2}
L_2
&=&
\frac{a_1}{2}\dot{Q}^2
+\frac{a_2}{2} Q^2
+\frac{a_3}{2} R^2
+\frac{b_1}{2}\dot{q}^2
+b_2 \dot{Q} \dot{q}
-V(\phi,q)
\nonumber\\
&+&
\xi 
\left(\dot{\phi}-R\right)
+\lambda
\left(\dot{R}-Q\right).
\ee
It is straightforward to recover the theory \eqref{3rd2}
after eliminating the auxiliary fields $R$ and $Q$
by varying the Lagrangian \eqref{3rd2} with respect to 
the Lagrange multipliers $\lambda$ and $\xi$.

We regard $\phi$, $R$, $Q$, and $q$
as the dynamical variables,
and define the conjugate momenta by
\be
&&
P_Q
:= \frac{\partial L_2}
                 {\partial \dot{Q}}
=a_1\dot{Q}+b_2 \dot{q},
\\
&&
P_q:=
\frac{\partial L_2}
                 {\partial \dot{q}}
= b_1\dot{q}+b_2 \dot{Q},
\\
&&
P_R
:=
\frac{\partial L_2}
                 {\partial \dot{R}}
= \lambda,
\\
&&
P_\phi
:=
\frac{\partial L_2}
                 {\partial \dot{\phi}}
=\xi.
\ee
First, we consider the nondegenerate case
\be
\label{dege}
\frac{\partial^2 L_2}{\partial \dot{Q}^2}
\frac{\partial^2 L_2}{\partial \dot{q}^2}
-
\left(\frac{\partial^2 L_2}{\partial \dot{Q}\partial\dot{q}}\right)^2
=
b_1a_1-b_2^2
\neq 0.
\ee
In this case, 
by rewriting $\dot{Q}$ and $\dot{q}$ in terms of $P_Q$ and $P_q$, 
\be
\dot{q}=
\frac{a_1 P_q-b_2 P_Q}
        {b_1a_1-b_2^2},
\qquad
\dot{Q}
=
\frac{-b_2 P_q+b_1 P_Q}
        {b_1 a_1-b_2^2},
\ee
we obtain the Hamiltonian 
\be
\label{ham_ng}
H
&:=&
  P_Q\dot{Q}
+P_R\dot{R}
+P_\phi\dot{\phi}
+P_q\dot{q}
-L_2
\nonumber\\
&=&
\frac{1}{2(b_1 a_1-b_2^2)}
\left(
a_1 P_q^2
-2b_2 P_q P_Q
+b_1 P_Q^2
\right)
\nonumber\\
&+&
P_R Q
+P_\phi R
-\frac{a_2}{2}Q^2
-\frac{a_3}{2}R^2
+ V(\phi,q).
\ee
Thus, the Hamiltonian \eqref{ham_ng}
is not bounded from below,
because of the linear dependence on 
the momenta $P_\phi$ and $P_R$.
In other words,
the theory \eqref{3rd1}
contains
the two Ostrogradsky ghosts.

Second, we consider the degenerate case 
\be
\label{dege}
\frac{\partial^2 L_2}{\partial \dot{Q}^2}
\frac{\partial^2 L_2}{\partial \dot{q}^2}
-
\left(\frac{\partial^2 L_2}{\partial \dot{Q}\dot{q}}\right)^2
=
b_1 a_1-b_2^2
=0,
\ee
under which $P_Q$ and $P_q$
satisfy
\be
P_q-\frac{b_2}{a_1}P_Q
= 0.
\ee
Regarding
\be
\label{x1}
X_1:=
P_q-\frac{b_2}{a_1}P_Q\approx 0,
\ee
as the primary constraint,
the total Hamiltonian can be defined as  
\be
\label{toth}
{\tilde H}
&:=&
H+\mu X_1
=
 \left(
  P_Q\dot{Q}
+P_R\dot{R}
+P_\phi\dot{\phi}
+P_q\dot{q}
-L_2\right)
+\mu X_1
\nonumber\\
&=&
\frac{a_1}{2b_2^2}P_q^2
+P_R Q
+P_\phi R
-\frac{a_2}{2}Q^2
-\frac{a_3}{2}R^2
+V\left(\phi,q\right)
\nonumber\\
&+&
\mu X_1.
\ee
The time evolution of the primary constraint $X_1$
then
generates the secondary constraint
\be
\label{x2}
X_2
&:=&
\dot{X}_1
=\{X_1,{\tilde H} \}
\nonumber\\
&=&
-g_2\phi-g_3 q
+
\frac{b_2}{a_1}  (P_R-a_2 Q)
\approx
0,
\ee
where we define the Poisson bracket,
\be
\{U_1, U_2\} 
&:= &
\left(
\frac{\partial U_1}
        {\partial \phi}
\frac{\partial U_2}
       {\partial P_\phi}
-
\frac{\partial U_1}
        {\partial P_\phi}
\frac{\partial U_2}
       {\partial \phi}
\right)
\nonumber\\
&+&
\left(
\frac{\partial U_1}
        {\partial R}
\frac{\partial U_2}
       {\partial P_R}
-
\frac{\partial U_1}
        {\partial P_R}
\frac{\partial U_2}
        {\partial R}
\right)
\nonumber\\
&+&
\left(
\frac{\partial U_1}
       {\partial Q}
\frac{\partial U_2}
       {\partial P_Q}
-
\frac{\partial U_1}
        {\partial P_Q}
\frac{\partial U_2}
        {\partial Q}
\right)
\nonumber\\
&+&
\left(
\frac{\partial U_1}
        {\partial q}
\frac{\partial U_2}
        {\partial P_q}
-
\frac{\partial U_1}
        {\partial P_q}
\frac{\partial U_2}
        {\partial q}
\right),
\ee
which relates $P_R$ to the other phase space variables
and eliminates the term linear in $P_R$ in the total Hamiltonian \eqref{toth}.

There is still the other term linear in $P_\phi$ in the total Hamiltonian \eqref{toth}.
We note that 
no further constraint is generated if $\{X_2,X_1\}\neq 0$,
since the time evolution of $X_2$ fixes 
the Lagrange multiplier $\mu$.
In order to obtain enough constraints,
we have to impose
\be
\{
X_1,X_2
\}
=
g_3
-
\frac{b_2^2}{a_1^2}a_2
\approx
0.
\ee
The time evolution of
the secondary constraint $X_2$ 
provides the tertiary constraint 
\be
\label{x3}
X_3
&:=&
\dot{X_2}
=\{X_2, \tilde{H}\}
\nonumber\\
&=&
-g_2 R
-\frac{a_2}{a_1}P_q
-\frac{b_2}{a_1}
\left(
P_\phi
-a_3 R
\right)
\approx 
0,
\ee
which relates $P_\phi$ to the other phase space variables
and eliminates the term linear in $P_\phi$
in the Hamiltonian \eqref{toth}.
Since $\{X_3,X_1\}=0$,
the time evolution of $X_3$ provides the quaternary condition
\be
\label{x4}
X_4
&:= &
\dot{X}_3
=
\{
X_3,\tilde{H}
\}
\nonumber\\
&=&
 \frac{g_2b_2 q}{a_1}
+\frac{a_2^2b_2^2 q}{a_1^3}
-g_2 Q
\nonumber\\
&+&
\frac{a_3 b_2 Q}{a_1}
+\frac{a_2 g_2 \phi}{a_1}
+\frac{g_1 b_2 \phi}{a_1}
\approx 0.
\ee
Since $\{X_4,X_1\}\neq 0$, 
the time evolution of $X_4$,
$\dot{X_4}=\{X_4,\tilde{H}\}\approx 0$,
fixes the Lagrange multiplier $\mu$
and no further constraint is generated.

We note that 
all the constraints  $X_i \approx 0$ ($i=1,2,3,4$),
\eqref{x1}, \eqref{x2}, \eqref{x3}, \eqref{x4},
are the second-class ones, 
since
\be
&&
   \{X_1, X_2\}
= \{X_1,X_3\}
=0,
\nonumber\\
&&
\{X_1,X_4\}
=
-\{X_2,X_3\}
=\frac{b_2}{a_1^3}
\left(
-2a_1^2 g_2-a_2^2b_2+a_1a_3 b_2
\right),
\nonumber\\
&&
\{X_2,X_4\}=0,
\nonumber\\
&&
\{X_3,X_4\}
=\frac{b_2}{a_1^4}
\left(
a_2^3b_2
+a_1^2\left(2a_2 g_2+g_1b_2\right)
\right).
\ee
Starting from the $8$-dimensional phase space $(\phi, R,Q,q,P_\phi,R_R,P_Q,P_q)$,
the $4$ second-class constraints
leave $4(=2\times 2)$ independent variables in the phase space,
namely, $2$ degrees of freedom,
and hence
all the Ostrogradsky ghosts are removed.

\section{The coefficients in Eqs. \eqref{diff1} and \eqref{diff2}}
\label{app_b}

The coefficients in Eq. \eqref{diff1} are given by 
\begin{eqnarray}
\label{q1}
\Q_1 (t)
&:=&
-\frac{a^2}{2}\dot{\phi}_0F_2
-\frac{3a^2}{2}\dot{\phi}_0 F_{0,\Box\phi}
+a \dot{a} \dot{\phi}_0^2 F_5
\nonumber\\
&+&
3a\dot{a}\dot{\phi}_0^2 F_{0,\W}
+\frac{3a\dot{a}}{2}\dot{\phi}_0^2 F_{2,\Box\phi}
-3\dot{a}^2\dot{\phi}_0^3
F_{2,\W}
\nonumber\\
&-&
\frac{3\dot{a}^2}{2}\dot{\phi}_0^3
F_{5,\Box\phi}
+\frac{3 \dot{a}^3} {a}  \dot{\phi}_0^4
F_{5,\W},
\end{eqnarray}
\begin{eqnarray}
\Q_2 (t)
&:=&
-a^2 F_{0,\X}
+2\dot{a}^2F_5 
+6 \dot{a}^2 F_{0,\W}
+3\dot{a}^2F_{2,\Box\phi}
\nonumber\\
&-&
\frac{a\dot{a} F_2}
          {\dot{\phi}_0}
-\frac{3a \dot{a}  F_{0,\Box\phi}}{\dot{\phi}_0}
+a\dot{a} \dot{\phi}_0 F_{2,\X}
\nonumber\\
&-&
\frac{6\dot{a}^3 \dot{\phi}_0 F_{2,\W}}
          {a}
-\frac{3\dot{a}^3\dot{\phi}_0 F_{5,\Box\phi} }{a}
\nonumber\\
&-&
\dot{a}^2\dot{\phi}_0^2F_{5,\X}
+\frac{6\dot{a}^4\dot{\phi}_0^2 F_{5,\W} }{a^2}
+2a^2\ddot{\phi}_0F_{0,\Y} 
\nonumber\\
&-&
2a \dot{a}\dot{\phi}_0\ddot{\phi}_0 F_{2,\Y}
+2\dot{a}^2\dot{\phi}_0^2\ddot{\phi}_0 F_{5,\Y} 
-4a^2 \ddot{\phi}_0^2F_{0,\Z}
\nonumber\\
&+&
4a \dot{a} \dot{\phi}_0 \ddot{\phi}_0^2F_{2,\Z}
-4 \dot{a}^2\dot{\phi}_0^2\ddot{\phi}_0^2F_{5,\Z},
\end{eqnarray}
\begin{eqnarray}
\Q_3 (t)
&:=&
\frac{a\dot{a}}{2} F_{2,\Box\phi}
-\frac{a^2}{2\dot{\phi}_0}F_{0,\Box\phi}
+a^2\dot{\phi}_0 F_{0,\Y}
-\frac{\dot{a}^2\dot{\phi}_0}{2}F_{5,\Box\phi}
\nonumber\\
&-&a \dot{a}\dot{\phi}_0^2 F_{2,\Y}
+\dot{a}^2\dot{\phi}_0^3 F_{5,\Y}
-
a\dot{a}\ddot{\phi}_0 
F_{2,\W}
\nonumber\\
&+&
\frac{a^2\ddot{\phi}_0}
       {\dot{\phi}_0}
F_{0,\W}
-
4a^2 \dot{\phi}_0\ddot{\phi}_0
F_{0,\Z}
+
\dot{a}^2\dot{\phi}_0
\ddot{\phi}_0
F_{5,\W}
\nonumber\\
&+&
4a\dot{a}\dot{\phi}_0^2\ddot{\phi}_0 F_{2,\Z} 
-4\dot{a}^2\dot{\phi}_0^3\ddot{\phi}_0 F_{5,\Z},
\end{eqnarray}
\begin{eqnarray}
\label{q4}
\Q_4(t)
&:=&
-\frac{F_{0,\Box\phi}}{2}
+\frac{\dot{a}\dot{\phi}_0F_{0,\W} }{a}
+\frac{\dot{a}\dot{\phi}_0F_{2,\Box\phi} }
         {2a}
\nonumber\\
&-&\frac{\dot{a}^2\dot{\phi}_0^2F_{2,\W} } 
         {a^2}
-\frac{\dot{a}^2\dot{\phi}_0^2 F_{5,\Box\phi}}
         {2a^2}
+\frac{\dot{a}^3\dot{\phi}_0^3F_{5,\W}} {a^3}.
\end{eqnarray}

The coefficients in Eq. \eqref{diff2} are given by 
\begin{widetext}
\begin{eqnarray}
\label{a1}
\alpha_1(t)
&:=&
F_0
+\frac{3\dot{a}\dot{\phi}_0}{a}
   F_{0,\Box\phi}
+  \dot{\phi}_0^2
   F_{0,\X}
-\frac{6\dot{a}^2\dot{\phi}_0^2}{a^2}
   F_{0,\W}
-\frac{3\dot{a}\dot{\phi}_0^3}
         {a}
   F_{1,\Box\phi}
-\dot{\phi}_0^4 F_{1,\X}
+\frac{6\dot{a}^2 \dot{\phi}_0^4}{a^2}
   F_{1,W}
-\ddot{\phi}_0 
   F_2
-\frac{3\dot{a}\dot{\phi}_0\ddot{\phi}_0}{a} 
   F_{2,\Box\phi}
\nonumber\\
&&
-2\dot{\phi}_0^2\ddot{\phi}_0 F_{0,\Y}
-\dot{\phi}_0^2\ddot{\phi}_0 F_{2,\X}
+\frac{6\dot{a}^2\dot{\phi}_0^2\ddot{\phi}_0}{a^2}
   F_{2,\W}
+\frac{6\dot{a}\dot{\phi}_0^3\ddot{\phi}_0}{a^2}
   F_{3,\Box\phi}
+2\dot{\phi}_0^4 \ddot{\phi}_0
  F_{1,\Y}
+2\dot{\phi}_0^4\ddot{\phi}_0
  F_{3,\X}
\nonumber\\
&&
-\frac{12\dot{a}^2\dot{\phi}_0^4\ddot{\phi}_0}{a^2}
  F_{3,\W}
+\ddot{\phi}_0^2 F_5
+\frac{3\dot{a} \dot{\phi}_0\ddot{\phi}_0^2}{a}
  F_{5,\Box\phi}
+4\dot{\phi}_0^2\ddot{\phi}_0^2 F_{0,\Z}
+2\dot{\phi}_0^2\ddot{\phi}_0^2 F_{2,\Y}
+\dot{\phi}_0^2\ddot{\phi}_0^2 F_{5,\X}
\nonumber\\
&&
-\frac{6\dot{a}^2\dot{\phi}_0^2\ddot{\phi}_0^2}
          {a^2}
   F_{5,\W}
-\frac{12\dot{a} \dot{\phi}_0^3\ddot{\phi}_0^2}{a}
   F_{4,\Box\phi} 
-4\dot{\phi}_0^4\ddot{\phi}_0^2 
  F_{1,\Z}
-4\dot{\phi}_0^4\ddot{\phi}_0^2 
  F_{3,\Y}
-4\dot{\phi}_0^4\ddot{\phi}_0^2 
  F_{4,\X}
+\frac{24\dot{a}^2\dot{\phi}_0^4\ddot{\phi}_0^2}
         {a^2}
  F_{4,\W}
\nonumber\\
&&
-4\dot{\phi}_0^2\ddot{\phi}_0^3 F_{2,\Z}
-2\dot{\phi}_0^2\ddot{\phi}_0^3 F_{5,\Y}
+8 \dot{\phi}_0^4\ddot{\phi}_0^3F_{3,\Z}
+8\dot{\phi}_0^4\ddot{\phi}_0^3 F_{4,\Y}
+4\dot{\phi}_0^2\ddot{\phi}_0^4 F_{5,\Z}
-16\dot{\phi}_0^4\ddot{\phi}_0^4 F_{4,\Z},
\end{eqnarray}
\begin{eqnarray}
\alpha_2(t)
&:=&
 \frac{\dot{\phi}_0}{2}
         F_2
+\frac{\dot{\phi}_0}{2}
         F_{0,\Box\phi}
-\dot{\phi}_0^3 
   F_3
-\frac{\dot{\phi}_0^3}{2}
   F_{1,\Box\phi}
-\dot{\phi}_0^3 F_{0,\Y}
+\dot{\phi}_0^5 F_{1,\Y}
-\dot{\phi}_0\ddot{\phi}_0 F_{5}
-\dot{\phi}_0\ddot{\phi}_0 F_{0,\W}
-\frac{1}{2}\dot{\phi}_0\ddot{\phi}_0 F_{2,\Box\phi}
\nonumber\\
&&
+4\dot{\phi}_0^3\ddot{\phi}_0 
  F_4
+\dot{\phi}_0^3\ddot{\phi}_0 
 F_{1,\W}
+\dot{\phi}_0^3\ddot{\phi}_0
 F_{3,\Box\phi}
+4\dot{\phi}_0^3\ddot{\phi}_0 
  F_{0,\Z}
+\dot{\phi}_0^3\ddot{\phi}_0 
  F_{2,\Y}
-4\dot{\phi}_0^5\ddot{\phi}_0 
  F_{1,\Z}
-2\dot{\phi}_0^5\ddot{\phi}_0 
  F_{3,\Y}
+\dot{\phi}_0\ddot{\phi}_0^2
   F_{2,\W}
\nonumber\\
&&
+\frac{\dot{\phi}_0\ddot{\phi}_0^2}{2}
   F_{5,\Box\phi}
-2\dot{\phi}_0^3\ddot{\phi}_0^2
   F_{3,\W}
-2\dot{\phi}_0^3\ddot{\phi}_0^2
   F_{4,\Box\phi}
-4\dot{\phi}_0^3\ddot{\phi}_0^2
   F_{2,\Z}
-\dot{\phi}_0^3\ddot{\phi}_0^2
   F_{5,\Y}
+8\dot{\phi}_0^5\ddot{\phi}_0^2
   F_{3,\Z}
+4\dot{\phi}_0^5\ddot{\phi}_0^2
   F_{4,\Y}
-\dot{\phi}_0\ddot{\phi}_0^3
   F_{5,\W}
\nonumber\\
&&
+4\dot{\phi}_0^3\ddot{\phi}_0^3
   F_{4,\W}
+4\dot{\phi}_0^3\ddot{\phi}_0^3
   F_{5,\Z}
-16\dot{\phi}_0^5\ddot{\phi}_0^3
  F_{4,\Z},
\end{eqnarray}
\begin{eqnarray}
\alpha_3(t)
&:=&
 \frac{3\dot{\phi}_0^3}{2}
         F_{0,\Box\phi}
-\frac{3\dot{a}\dot{\phi}_0^4}{a}
          F_{0,\W}
-\frac{3\dot{\phi}_0^5}{2}
         F_{1,\Box\phi}
+\frac{3\dot{a}\dot{\phi}_0^6}{a}
          F_{1,\Box\phi}
-\frac{3\dot{\phi}_0^3\ddot{\phi}_0}{2}
         F_{2,\Box\phi}
+\frac{3\dot{a}\dot{\phi}_0^4\ddot{\phi}_0}{a}
          F_{2,\W}
+3\dot{\phi}_0^5\ddot{\phi}_0
         F_{3,\Box\phi}
-\frac{6\dot{a}\dot{\phi}_0^4\ddot{\phi}_0}{a}
          F_{3,\W}
\nonumber\\
&&
+\frac{3\dot{\phi}_0^3\ddot{\phi}_0^2}{2}
         F_{5,\Box\phi}
-\frac{3\dot{a}\dot{\phi}_0^4\ddot{\phi}_0^2}{a}
          F_{5,\W}
-6\dot{\phi}_0^5\ddot{\phi}_0^2
         F_{4,\Box\phi}
+\frac{12\dot{a}\dot{\phi}_0^6\ddot{\phi}_0^2}{a}
         F_{4,\W},
\end{eqnarray}
\begin{eqnarray}
\label{a4}
\alpha_4(t)
&:=&
\frac{\dot{\phi}_0^2}{2a^2}
  F_{0,\Box\phi}
-\frac{\dot{a}\dot{\phi}_0^3}{a^3}
  F_{0,\W}
-\frac{\dot{\phi}_0^4}{2a^2}
  F_{1,\Box\phi}
+\frac{\dot{a}\dot{\phi}_0^5}{a^3}
  F_{1,\W}
-\frac{\dot{\phi}_0^2\ddot{\phi}_0}{2a^2}
  F_{2,\Box\phi}
+\frac{\dot{a}\dot{\phi}_0^3\ddot{\phi}_0}{a^3}
  F_{2,\W}
+\frac{\dot{\phi}_0^4\ddot{\phi}_0}{a^2}
  F_{3,\Box\phi}
-\frac{2\dot{a}\dot{\phi}_0^5\ddot{\phi}_0}{a^3}
  F_{3,\W}
\nonumber\\
&&
+\frac{\dot{\phi}_0^2\ddot{\phi}_0^2}{2a^2}
  F_{5,\Box\phi}
-\frac{\dot{a}\dot{\phi}_0^3\ddot{\phi}_0^2}{a^3}
  F_{5,\W}
-\frac{2\dot{\phi}_0^4\ddot{\phi}_0^2}{a^2}
  F_{4,\Box\phi}
+\frac{4\dot{a}\dot{\phi}_0^5\ddot{\phi}_0^2}{a^3}
  F_{4,\W}
\end{eqnarray}
\end{widetext}

\bibliography{disformal_refs}
\end{document}